\documentclass[]{jpsj3}
\usepackage{txfonts}
\usepackage{here}
\usepackage[dvipdfm]{}
\usepackage{def}
\usepackage{bm}

\title{Quasi-localized Impurity State in Doped Topological Crystalline Insulator Sn$_{0.9}$In$_{0.1}$Te Probed by $^{125}$Te-NMR }

\author{
\name{Satoki \surname{Maeda}}$^1$,
\name{Shota \surname{Katsube}}$^1$,
\name{Guo-qing \surname{Zheng}}$^{1,2}$
}

\inst{
$^1$Department of Physics, Okayama University, Okayama 700-8530, Japan\\
$^2$Institute of Physics, Chinese Academy of Sciences and Beijing National Laboratory for Condensed Matter Physics, Beijing 100190, China\\
} 

\abst{
The In-doped topological crystalline insulator Sn$_{1-x}$In$_x$Te is a promising candidate for a topological superconductor, where it is theoretically suggested that In creates an impurity state responsible for superconductivity.
We synthesized high purity Sn$_{1-x}$In$_x$Te samples and performed $^{125}$Te-nuclear magnetic resonance (NMR) measurements.
The NMR spectra under a magnetic field of $H_0$ = 5 T show a broadening characteristic due to a localized impurity state.
The spin-lattice relaxation rate ($1/T_1$) divided by temperature shows a Curie-Weiss like temperature-dependence under $H_0$ = 0.1 T but is temperature-independent under $H_0$ = 5 T.
These results indicate the existence of quasi-localized impurity states due to In doping.
}

\begin{document}
\maketitle

\section{Introduction}
A topological insulator (TI) is a material in which the bulk is insulating but the surface contains metallic state due to non-zero topological invariants of the bulk band structure\cite{Hasan,Qi,Ando}.
SnTe with NaCl-type crystal structure is a topological crystalline insulator (TCI)\cite{Hsieh,Tanaka}.
A TCI requires certain symmetries in crystal structure such as mirror symmetry while a TI requires time-reversal symmetry. 
It has been proposed that the superconductivity in In-doped SnTe may be topological.\cite{Sasaki}
Topological superconductivity is anologous to TI in that the superconducting gap function has a nontrivial topological invariant.
The best studied candidate of topological superconductors is a doped TI, Cu$_x$Bi$_2$Se$_3$,\cite{Hor} in which a spin-triplet, odd-parity superconducting state was recently established\cite{Matano}.

The superconducting properties of doped SnTe has large dopant-dependence.
When the Sn site is doped with Ag or vacancy, the Fermi level ($E_{\rm F}$) is simply lowered, and superconductivity appears at low temperatures (transition temperature $T_{\rm c}\sim 0.1$ K with a hole concentration of $p = 10^{21}$cm$^{-3}$)\cite{Mathur}.
On the other hand, when the Sn site is doped with In, superconductivity appears at higher temperatures ($T_{\rm c} \sim 1$ K with $p = 10^{21}$cm$^{-3}$).\cite{Erickson1}
The existence of impurity states (in-gap states) bound to In was suggested from the results of a reduction of carrier mobility and the pinning of the $E_{\rm F}$ due to In-doping.\cite{Kaidanov,Bushmarina,Nemov}
Based on this, Shelankov proposed that a strong interaction between the impurity state and phonon is responsible for the enhancement of the superconducting transition temperature.\cite{Shelankov}
More recently, it is suggested by an first principle calculation that the impurity state is composed primarily of In $5s$ and Te $5p$ orbitals .\cite{Haldo}

Such a bound state often contributes to magnetism.
For example, in P-doped Si with P-concentrations $N_{\rm P} < 6.5\times 10^{18} \rm{cm}^{-3}$, the magnetic susceptibility ($\chi$) shows a Curie-Weiss like temperature ($T$)-dependence.\cite{Kobayashi,Ikehata}.
Several nuclear magnetic resonance (NMR) studies have also been reported.
The magnetic field swept $^{29}$Si-NMR (nuclear spin $I=1/2$) spectrum shows a characteristic broadening with a tail on the low field side because the wave function of the localized impurity state creates an inhomogeneous magnetic field.
The spin-lattice relaxation rate ($1/T_1$) divided by $T$ shows a Curie-Weiss like $T$-dependence at low fields due to the magnetization of impurity states but is $T$-independent under high magnetic fields\cite{Kobayashi}.

Sn$_{1-x}$In$_x$Te has four Fermi surfaces centered at L points of the fcc lattice while Cu$_x$Bi$_2$Se$_3$ has one Fermi surface centered at $\Gamma$ point of the rhombohedral lattice, but each Fermi surface in Sn$_{1-x}$In$_x$Te is essentially equivalent to that of Cu$_x$Bi$_2$Se$_3$\cite{Sasaki}.
Point-contact spectroscopy found a zero-bias conductance peak which was taken as a signature of unconventional superconductivity\cite{Sasaki}.
On the other hand, specific heat have revealed fully gapped superconductivity\cite{Mario}.
By combining these results, a spin-triplet state with an isotropic gap was suggested\cite{Hashimoto}.
Therefore, NMR measurements which can reveal spin symmetry as well as the parity of the gap function in the superconducting state is desired.
As a first step toward a full understanding of this material, we aim to observe the impurity state by NMR.
In this paper, we report the synthesis of Sn$_{1-x}$In$_x$Te ($x$ = 0 and 0.1) and $^{125}$Te-NMR ($I=1/2$) measurements.

\section{Experimental}
Polycrystalline samples of SnTe and Sn$_{0.9}$In$_{0.1}$Te were synthesized by a sintering method.
The required amounts of Sn, In, and Te were pre-reacted in evacuated quartz tubes at 1000$^\circ$C for 8 hours.
The resultant materials were powdered, pressed into pellets, and sintered in evacuated quartz tubes at 400$^\circ$C for 2 days.
In order to avoid the influence of dilute magnetic impurities such as Fe , high purity starting materials (Sn: 99.9999\%, In: 99.9999\%, and Te: 99.9999\%) were used.
For powder x-ray diffraction (XRD), ac susceptibility, and NMR measurements, a part of the pellet was powdered.
The samples were characterized by powder XRD using Rigaku RINT-TTR III with CuK$\alpha$ radiation.
No impurity peaks were observed in the powder XRD pattern.
The $T_{\rm c}$ was determined by measuring the inductance of a coil filled with a sample which is a typical setup for NMR measurements.
NMR measurements were carried out by using a phase-coherent spectrometer.
The NMR spectrum was obtained by integrating the spin echo intensity by changing the resonance frequency ($f$) at the fixed magnetic field of 5 T.
The spin-lattice relaxation time ($T_1$) was measured by using a single saturating pulse, and determined by fitting the recovery curve of the nuclear magnetization to the exponential function: $(M_0-M(t))/M_0 = \exp(-t/T_1)$, where $M_0$ and $M(t)$ are the nuclear magnetization in the thermal equilibrium and at a time $t$ after the saturating pulse.
The recovery curves in the whole temperature and magnetic field ranges were well fitted by the single exponential function.

\section{Results and Discussion}
\begin{figure}[H]
\begin{center}\includegraphics[clip,width=75mm]{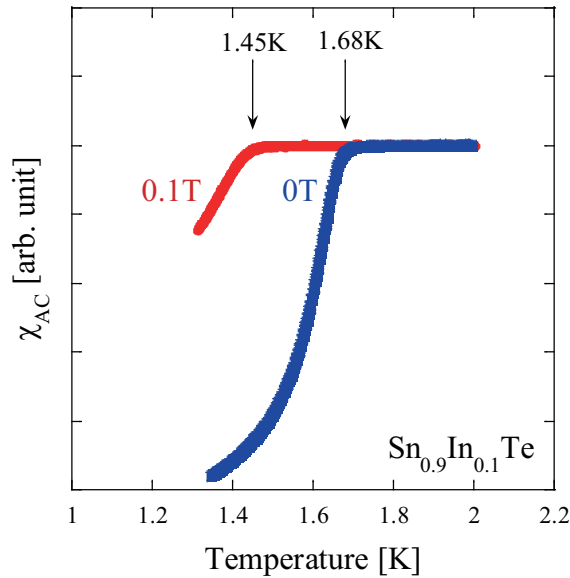}\end{center}
\caption{(Color online) Temperature-dependence of the ac susceptibility for Sn$_{0.9}$In$_{0.1}$Te under static magnetic fields of $H_0$ = 0 T and 0.1 T, respectively.
The solid arrows indicate $T_{\rm c}$.}
\label{chi}
\end{figure}

Figure \ref{chi} shows the $T$-dependence of the ac susceptibility for Sn$_{0.9}$In$_{0.1}$Te, which showed superconductivity at 1.68 K under $H_0 = 0$ T and at 1.45 K under $H_0 = 0.1$ T.
To obtain information on superconductivity of Sn$_{0.9}$In$_{0.1}$Te by NMR, the applied magnetic field must be 0.1 T or less, owing to the small upper critical field $H_{\rm c2}$.

\begin{figure}[H]
\begin{center}\includegraphics[clip,width=75mm]{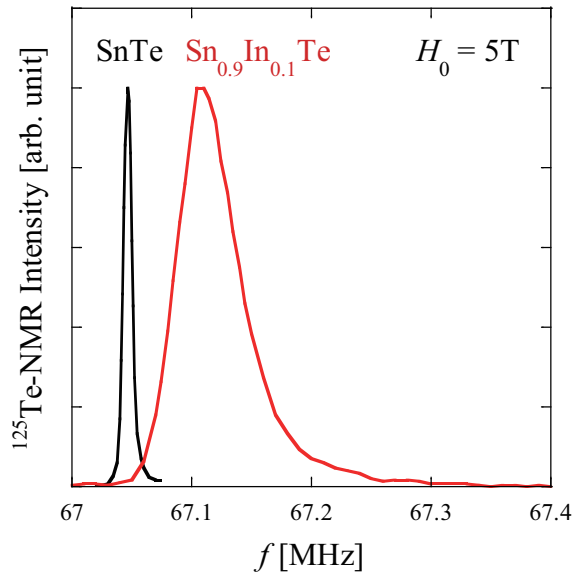}\end{center}
\caption{(Color online) $^{125}$Te-NMR spectra for SnTe at 3 K and Sn$_{0.9}$In$_{0.1}$Te at 1.55 K under $H_0$ = 5 T.}
\label{Spectra1}
\end{figure}

\begin{figure}[H]
\begin{center}\includegraphics[clip,width=75mm]{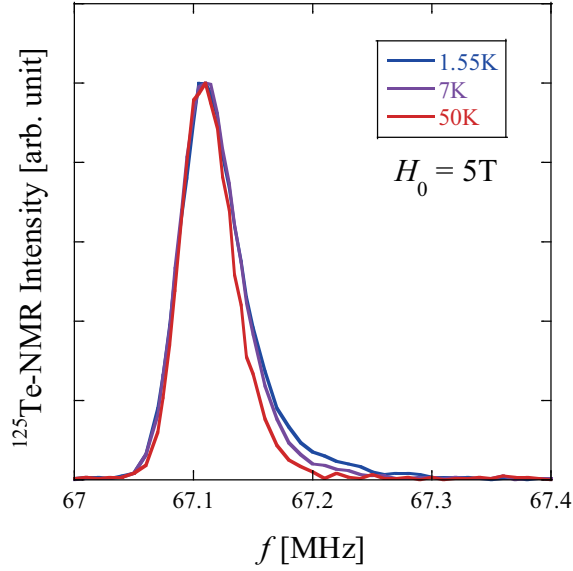}\end{center}
\caption{(Color online) $^{125}$Te-NMR spectra for Sn$_{0.9}$In$_{0.1}$Te at 1.55 K, 7 K, and 50 K under $H_0$ = 5 T.}
\label{Spectra2}
\end{figure}

Figure \ref{Spectra1} shows the $^{125}$Te-NMR spectra under $H_0 = 5$ T.
The $^{125}$Te Knight shift ($K$) in NaCl type crystal structure is isotropic because the Te site is at the center of the regular tetrahedron of Sn/In.
Although the non-doped SnTe shows a small distortion at about 100 K\cite{Erickson1}, the sharp spectrum of SnTe indicates the anisotropy is negligibly small.
Such distortion is completely suppressed in Sn$_{0.9}$In$_{0.1}$Te\cite{Erickson1}.
The spectrum of Sn$_{0.9}$In$_{0.1}$Te has a large tail on the high frequency side resembling that observed in P-doped Si\cite{Kobayashi}.
The $K$ is expressed as $K = K_{\rm s} + K_{\rm orb}$, where $K_{\rm s}$ is the spin part, and $K_{\rm orb}$ is the orbital part.
The difference in peak position between SnTe and Sn$_{0.9}$In$_{0.1}$Te is due to the $K_{\rm s}$ that is sensitive to the carrier concentration.
Figure \ref{Spectra2} shows the spectra of Sn$_{0.9}$In$_{0.1}$Te at three different temperatures.
As the temperature is lowered, the tail on the high frequency side grew while the peak position was unchanged, which means that the spin polarization of the impurity states developed.
This situation is clearly different from the case due to dilute magnetic impurities such as Fe or dense magnetic ions in the lattice, in which cases the NMR spectrum is symmetrically broadened.\cite{CuFe,Nd}

\begin{figure}[H]
\begin{center}\includegraphics[clip,width=80mm]{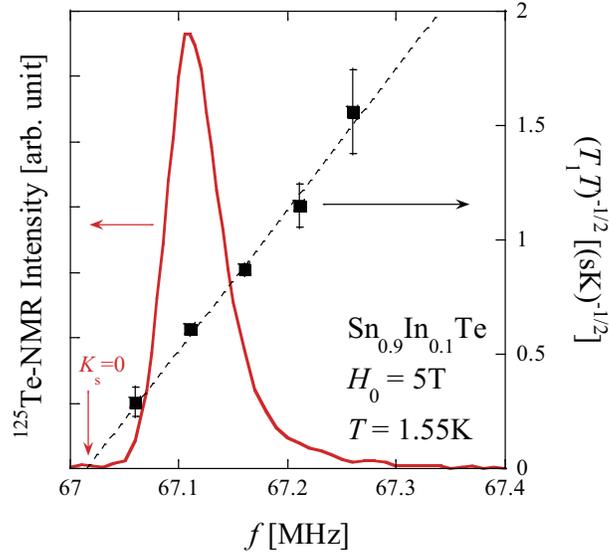}\end{center}
\caption{(Color online) Frequency-dependence of the $(T_1T)^{-1/2}$ for Sn$_{0.9}$In$_{0.1}$Te at 1.55 K under $H_0$ = 5 T. The vertical arrow indicates the position of $K_{\rm s}$ = 0.}
\label{T1f}
\end{figure}

Figure \ref{T1f} shows the $f$-dependence of the $\left( T_1T \right)^{-1/2}$ at 1.55 K under $H_0 = 5$ T.
The higher the $f$, the larger the $\left( T_1T \right)^{-1/2}$ became, which means that the $1/T_1$ at the tail is enhanced reflecting the large density of states due to the in-gap state.
The linear relation between $\left( T_1T \right)^{-1/2}$ and $f$ suggests that a Korringa relation $T_1\left( \bm r \right)TK_{\rm s}^2\left( \bm r \right) = a$ is satisfied locally, where $T_1\left( \bm r \right)$ is the $T_1$ at a position of $\bm r$, $K_s\left( \bm r \right)$ is the $K_{\rm s}$ at $\bm r$, and $a$ is a constant.
By extrapolating the linear relation $K$ vs $(1/T_1T)^{-1/2}$ to the origin where $(1/T_1T)^{-1/2} = 0$, the value of $K_s$ was obtained.
If the electron correlation is weak and the $s$-orbital electrons make the dominant contribution, the $a$ is calculated as, $a_{\rm s} = \frac{\hbar}{4\pi k_{\rm B}}\left( \frac{\gamma_{\rm e}}{\gamma_{\rm n}}\right)^2$, where $k_{\rm B}$ is the Boltzmann constant, $\gamma_{\rm e}$ is the gyromagnetic ratio of electrons, and $\gamma_{\rm n}$ is the nuclear gyromagnetic ratio.
We obtained $a = 2.2a_{\rm s}$ for Sn$_{0.9}$In$_{0.1}$Te.
The deviation form $a_{\rm s}$ is most likely due to the contribution of the Te $5p$ orbital as suggested by the first principle calculation\cite{Haldo}.
Similar inhomogeneity of the $T_1$ was also reported for P-doped Si\cite{Kobayashi}.

\begin{figure}[H]
\begin{center}\includegraphics[clip,width=80mm]{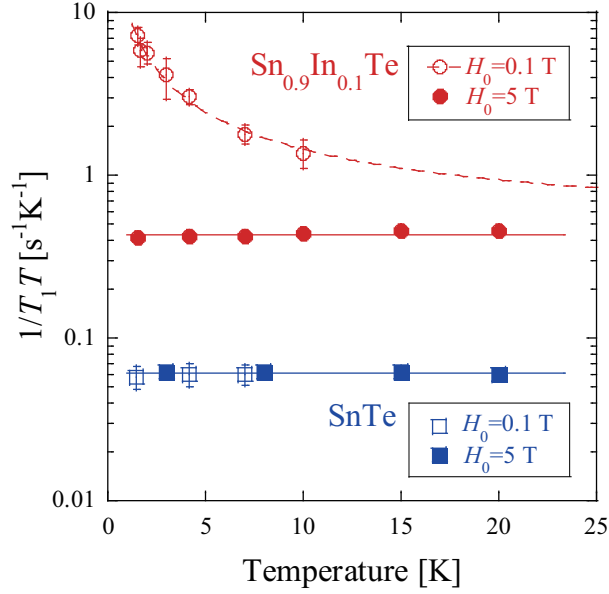}\end{center}
\caption{(Color online) Temperature-dependence of the $1/T_1T$ for SnTe (blue squares) and Sn$_{0.9}$In$_{0.1}$Te (red circles) under $H_0$ = 0.1 T (open markers) and 5 T (filled markers). The red dashed line represents the relation $1/T_1T = A + B/T$, with $A = 0.43$ [s$^{-1}$K$^{-1}$] and $B = 10.1$ [s$^{-1}$].}
\label{T1T}
\end{figure}

Figure \ref{T1T} shows the $T$- dependence of the $1/T_1T$ at different $H_0$ for SnTe and Sn$_{0.9}$In$_{0.1}$Te measured at the peak position.
The $1/T_1T$ of SnTe is $H_0$- and $T$-independent, indicating a conventional metallic state of SnTe and that the amount of magnetic impurities such as Fe is negligibly small.
It is well known that SnTe is actually metallic because of Sn-vacancy although an ideal SnTe is a semiconductor\cite{Erickson1}.
On the other hand, the $1/T_1T$ of Sn$_{0.9}$In$_{0.1}$Te shows a distinct Curie-Weiss like $T$-dependence under $H_0$ = 0.1 T but is $T$-independent under $H_0$ = 5 T.

According to Erickson $et\ al$., the $T$-dependence of the $\chi$ in Sn$_{1-x}$In$_x$Te is Pauli like rather than Curie-Weiss like.\cite{Erickson2}
Below, we explain why both NMR spectra and $1/T_1T$ in Sn$_{0.9}$In$_{0.1}$Te show the peculiar characteristics even though the $\chi$ is Pauli like, by referring to and partly modifying the discussion of P-doped Si given by Kobayashi $et\ al$.\cite{Kobayashi}
With high impurity concentration, the impurity states hybridize and create a narrow band with finite density of states at the $E_{\rm F}$.
The wave functions of the impurity states can be extended over many impurity sites.
The states near the $E_{\rm F}$ are spin polarized under an external field and give the Pauli like $\chi$.
However, because of the localized nature of the impurity states, an internal inhomogeneous magnetic field is created.
The $^{125}$Te nuclei within the wave functions of the polarized impurity states feel large magnetic fields and compose the tail on the high frequency side.
As the temperature is lowered, the tail extends to the high frequency side because the magnetization of the impurity states develops.
On the other hand, the $^{125}$Te nuclei without the wave function of the polarized impurity states compose the large area on the low frequency side and have small $T$-dependence. 


The $1/T_1T$ under high magnetic fields satisfies the Korringa relation locally.
On the other hand, under low fields, where the NMR spectrum broadening due to the impurity states is smaller than that due to the nuclear dipole interaction, a spin-diffusion process homogenizes the $1/T_1T$.
Kobayashi $et\ al$. qualitatively explained the $T$-dependence of the $1/T_1T$ in P-doped Si with $N_{\rm P} < 6.5\times 10^{18}$ cm$^{-3}$ but did not explain that in P-doped Si with $N_{\rm P} \geq 6.5\times 10^{18}$ cm$^{-3}$ which showed a behavior similar to the present results for Sn$_{0.9}$In$_{0.1}$Te.\cite{Kobayashi}

We thus propose that the $1/T_1T$ under low magnetic fields is obtained as the spatial average of $1/T_1(\bm{r})T$,
\begin{equation}
\frac{1}{T_1T} = \frac{1}{V} \int \frac{\mathrm{d}^3 \bm{r}}{T_1(\bm{r})T} = \frac{1}{Va} \int K_s^2 (\bm{r}) \ \mathrm{d}^3 \bm{r},
\end{equation}
where $V$ is the volume.
By assuming $\mu_{\rm B}H \ll  k_{\rm B}T$, the integral around the polarized impurity states is proportional to $T^{-1}$.
This is because the effective number of the polarized impurity state is determined by $k_{\rm B}T$, while $K_s(\bm{r})$ proportional to the spin polarization of the impurity states goes as $T^{-1}$.\cite{Kobayashi}
On the other hand, the integral away from the polarized impurity states is $T$-independent.
Therefore, the $1/T_1T$ is linear in $T^{-1}$, which explains the Curie-Weiss like $T$-dependence of $1/T_1T$ in Sn$_{0.9}$In$_{0.1}$Te, as well as in highly P-doped Si.


Generally, the $1/T_1T$ which can probe the parity of the superconducting gap function is a very important quantity.
However, a Curie-Weiss like $T$-dependence of $1/T_1T$ obscures the important information and is a serious obstacle.
This was encountered in the study of the electron-doped high-$T_{\rm c}$ cuprates Nd$_{1-x}$Ce$_x$CuO$_4$\cite{Nd}, and is also true for Cu$_x$Bi$_2$Se$_3$.
Nisson $et\ al$. reported a Curie-Weiss like $T$-dependence of $1/T_1T$ by $^{209}$Bi-NMR ($I = 9/2$) and suggested that it is due to the magnetism of Se-vacancies\cite{Nisson}.
Therefore, an efficient method for measuring $1/T_1$ for these systems should be worked out.

\section{Summary}
We synthesized high purity Sn$_{1-x}$In$_x$Te polycrystals and performed $^{125}$Te-NMR measurements.
The NMR spectra under $H_0 = 5$ T showed a broadening characteristic due to a localized impurity state.
The $1/T_1T$ showed a Curie-Weiss like $T$-dependence under $H_0 = 0.1$ T but was $T$-independent under $H_0 = 5$ T.
These results indicate the existence of a quasi-localized impurity states due to In-doping.
Since such state was proposed to be responsible for the superconductivity, our results serve to lay a foundation toward understanding the possible exotic superconducting state of this material.

\begin{acknowledgment}
We acknowledge partial support by MEXT Grant No. 15H05852 and JSPS grant No.16H0401618.
\end{acknowledgment}

\end{document}